\documentclass{jltp}

\usepackage{graphicx} 
\title{Anisotropic scaling in YBa$_2$Cu$_3$O$_7$ crystals with columnar defects}

\author{Alejandro V. Silhanek$^*$ and Leonardo Civale$^+$}

\address{$^*$Laboratorium voor Vaste-Stoffysica en Magnetisme, K. U. Leuven,\\ Celestijnenlaan 200 D, B-3001 Leuven, Belgium.\\$^+$Superconductivity Technology Center, MS K763, \\ Los Alamos National Laboratory, Los Alamos, NM 87545, USA}

\runninghead{A.~V.~Silhanek and L.~Civale}{Anisotropic scaling in YBa$_2$Cu$_3$O$_7$ crystals with columnar defects}

\begin{document}

\maketitle

\begin{abstract}
We explore the interplay between the mass anisotropy and the uniaxial pinning of tilted columnar defects (CD) in a YBa$_2$Cu$_3$O$_7$ single crystal. At high temperatures T and fields H a sharp peak in the irreversible magnetization $M_i$ at the direction of the tracks signals the presence of the CD. At low T such a peak is not observed, and the influence of the CD appears only as a perturbation to the angular dependence due to the anisotropy. We show that at low T and high H the uniaxial effects due to CD can be recovered by performing an anisotropic rescaling. As expected, this scaling fails at low H.

\end{abstract}

It is well established that correlated disorder in high temperature superconductors (HTSC) generates a variety of vortex structures depending on the orientation of the defects and the applied field $\bf H$.\cite{evidence} In the solid vortex phase, when the angle between $\bf H$ and the extended defects is small, vortices remain locked into the defects, whereas for larger angles staircase vortices develop.\cite{evidence,avila01} In principle, this picture applies with minor differences to any correlated pinning potential.\cite{blatter94} Recently we have confirmed experimentally the presence of a lock-in phase in YBa$_2$Cu$_3$O$_7$ crystals with columnar defects and shown that the combined effect of CD, twins and the intrinsic pinning produces several different types of zigzag vortex structures.\cite{review}

The crystalline anisotropy, originated in the layered structure of the HTSC and characterized by the mass anisotropy $\gamma = m_c/m_{ab} \gg 1$, also influences the pinning in various ways. First, a large $\gamma$ implies a small vortex line tension, thus resulting in a broad lock-in and staircases angular regimes. Second, the angle-dependent line tension tends to tilt the vortices towards the $ab$-planes, thus competing with the uniaxial pinning of the CD.

In this work we explore the interplay between mass anisotropy and uniaxial pinning produced by aligned CD in a YBa$_2$Cu$_3$O$_7$ single crystal. The CD were introduced by irradiation with 315 MeV Au$^{+23}$ ions at the TANDAR facility, at an angle $\theta_D = 32^\circ$ from the $c$ axis, with a dose-equivalent matching field $B_{\Phi}=$ 3 T. To that end we study the angular dependence of the irreversible dc-magnetization using a SQUID magnetometer with two sets of pick up coils that allows to determine two perpendicular components of the magnetization $\bf M$. The crystal was rotated around an axis perpendicular to both its normal (that coincides with the $c$ axis) and $\bf H$. From the width of isothermal hysteresis loops recorded for a set of angles $\theta$ ($\theta=0^\circ$ for ${\bf H} \parallel c$) we determine the modulus of the irreversible magnetization $M_i(T,H,\theta)$, which is proportional to the persistent current density $J$ according to the critical state model. Further experimental details can be found in ref.[\onlinecite{review}].

\begin{figure}[t]
\centerline{\includegraphics[width=4.5in]{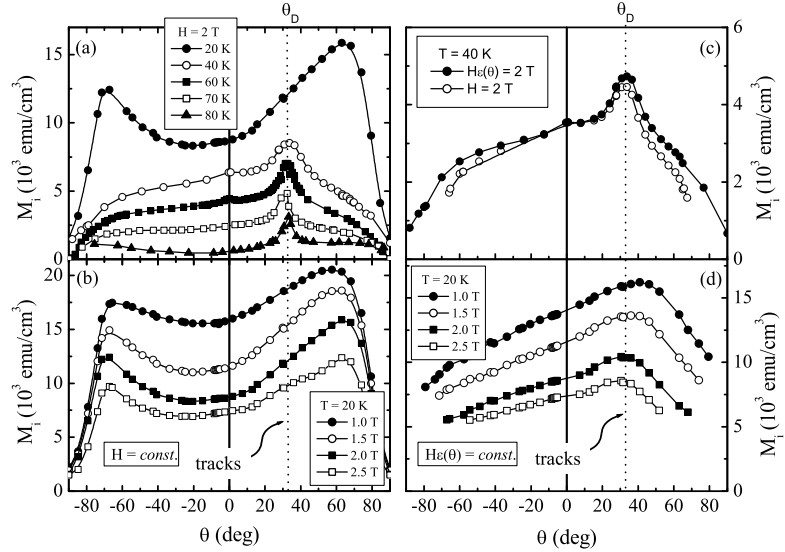}}
\caption{Modulus of the irreversible magnetization $M_i$ vs. angle $\theta$ for a YBa$_2$Cu$_3$O$_7$ crystal with CD, at (a) $H = 2$ T and various $T$; and (b) $T$ = 20 K and various $H$. For clarity, $M_i$ values are multiplied by factors 1.7; 3; 4 and 10 for $T$ = 40, 60, 70 and 80 K, respectively. (c) $M_i(\theta)$ at $T = 40$ K for $H = 2$ T and $H\varepsilon(\theta)=2$ T. (d) $M_i(\theta)$ at $T = 20$ K for $H\varepsilon(\theta)= const.$}  
\label{fig1}
\end{figure}

Figure \ref{fig1}(a) shows $M_i(\theta)$ for several temperatures at $H$ = 2 T. At high $T$ ($\geq 40$ K) the peak at the direction of the tracks $\theta_D$ (dashed line), signaling the uniaxial pinning of the CD, is clearly visible. At $T = 20$ K, in contrast, $M_i$ grows as $\bf H$ is tilted off the $c$ axis in both directions, and there is no CD's peak. The same qualitative behavior occurs for $T < 20$ K. Fig. \ref{fig1}(b) shows $M_i(\theta)$ at 20 K for several $H$. No hint of the CD's peak is observed for $H \le 2$ T, whereas a minor ``bump" at $\theta_D$ develops for $H = 2.5$ T. However, clear evidence of the uniaxial pinning of the tilted CD is visible at all $H$ as an asymmetry of $M_i(\theta)$ around $\theta = 0^\circ$: $M_i(\theta)$ is larger at $\theta > 0^\circ$ than at the symmetric angle $-\theta$. We note that the decrease of $M_i(\theta)$ as $\theta \rightarrow 90^\circ$ is due to the change in the geometrical factor and not to a decrease in $J_c$.

The increase of $M_i(\theta)$ at fixed $H$ as $\theta$ grows is characteristic of $J_c(\theta)$ in HTSC due to anisotropy.\cite{blatter94} The traditional way to account for anisotropy in a superconductor is to incorporate an effective-mass tensor $m_{ij}$ into the GL or London equations.\cite{kogan81} Alternatively, Blatter et al.\cite{blatter92} showed that the expressions valid for isotropic superconductors can be extended to the anisotropic case by performing a suitable rescaling of coordinates and the vector potential. They obtained an anisotropic scaling rule $Q(\theta,H,T,\xi,\lambda,\gamma) = s_Q \bar Q(\varepsilon(\theta)H,\gamma T,\xi,\lambda)$ for a desired quantity $Q$ for which the isotropic result $\bar Q$ is known. Here $\varepsilon(\theta)^2 = \cos^2\theta+\gamma^{-2} \sin^2\theta$ and $\xi$ and $\lambda$ refer to the values for ${\bf H} \parallel c$-axis. The scaling factor $s_Q = \gamma^{-1}$ for volumes, energies, actions and temperatures, and $s_Q = 1/\varepsilon(\theta)$ for magnetic fields. Thus, a field $H$ at an angle $\theta$ maps into a field of intensity $\tilde H =$ $H\varepsilon(\theta)$ and orientation $\tilde \theta = \arctan(\gamma^{-1} \tan \theta)$ in a ``fictitious" isotropic superconductor. Scalar disorder can be included in this description. It was shown\cite{blatter94} that $J_c(H,\theta)$ due to random defects depends on a single variable, $J_c(H,\theta) = J_c(\tilde H)$. In contrast, the uniaxial nature of the pinning of CD (or any correlated pinning) is not removed by this procedure.

We applied the scaling approach to the data of Fig. \ref{fig1}(b), using $\gamma = 7$. The result is shown in Fig. \ref{fig1}(d), where a large number of $M_i(H,\theta)$ data points were used to build curves of $M_i(\theta)$ at $\tilde H = const$. The contrast with Fig. \ref{fig1}(b) is apparent. Once the influence of the mass anisotropy was removed by the scaling, the peak centered at the tracks is recovered and the angular variation of the pinning due to the CD emerges clearly. 

\begin{figure}[t]
\centerline{\includegraphics[width=2.8in]{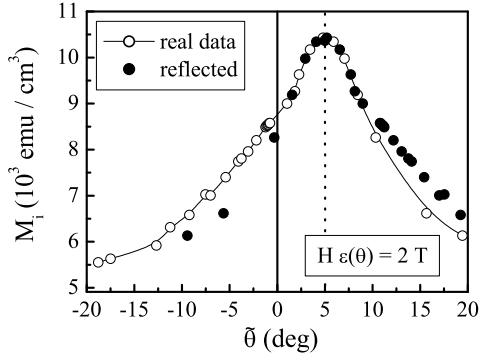}}
\caption{Modulus of irreversible magnetization $M_i$ vs $\tilde \theta$ at $T = 20$ K and $\tilde H = 2$ T, and its mirror reflection around $\tilde \theta_D = 5^\circ$ (dashed line)}
\label{fig2}
\end{figure}

The use of the scaling approach does not modify significantly the high $T$ results, because the $H = const.$ curves already exhibit a narrow peak at $\theta_D$, which is only slightly distorted by the transformation. This is seen in fig. \ref{fig1}(c), where we show the $M_i(H,\theta)$ curves at $T$ = 40 K for both constant $\tilde H$ = 2 T and constant $H$ = 2 T. The situation at higher $T$ and $H$ is similar. 

The curves in Fig. 1(d) are asymmetric with respect to $\theta_D$. The reason is that the correct angular variable in the fictitious isotropic superconductor is $\tilde \theta$ rather than $\theta$. In Fig. 2 we plot $M_i$ vs. $\tilde \theta$ at 20 K and $\tilde H = 2$ T, and its mirror image around $\tilde \theta_D = 5^\circ$. The peak is now almost symmetric. The small residual asymmetry is due to angular effects that are symmetric with respect to $\tilde \Theta = 0$, such as sample geometry and pinning by twin boundaries.



The conditions for the validity of the scaling approach, namely that both the spatial variations of the magnetic induction $\bf B$ and its misalignment with respect to $\bf H$ are negligible,\cite{blatter92} are not satisfied at low $H$. Indeed, the uniformity condition implies $\lambda \gg a_0=\sqrt{\Phi_0/B}$, which is equivalent to $H \gg$ 0.1 T. On top of that, we have shown\cite{NbSe} that $\bf B$ deviates from $\bf H$ for $H \le 0.02 H_{c2}$, where $H_{c2} \sim 1.6$[T/K]$(T_c-T)$ is the upper critical field. We thus estimate $H \sim 2$ T as a lower bound for the validity of the scaling. This value coincides with the field at which a slight deviation of the maximum with respect to $\theta_D$ in the $\tilde H = const.$ curves starts, strongly suggesting that this shift is a consequence of the breakdown of the scaling rule.

In summary, we have shown that the anisotropic scaling is an important tool to discriminate between uniaxial pinning by tilted CD and anisotropy effects in YBa$_2$Cu$_3$O$_7$ crystals at low temperatures.

We thank the Atomic Energy Commission of Argentina where the measurements were performed and CONICET of Argentina for financial support.

\end{document}